\documentclass[11pt]{article}
\usepackage[margin=1in]{geometry}
\usepackage[T1]{fontenc}
\usepackage{lmodern,microtype}
\usepackage{amsmath,amssymb,amsthm,mathtools}
\usepackage{tikz}
\usetikzlibrary{decorations.pathreplacing,arrows.meta}
\usepackage[section]{placeins}
\usepackage{float}
\usepackage[sorting=nyt,backend=bibtex,style=alphabetic]{biblatex}
\AtEveryBibitem{\ifkeyword{comparison-added}{\color{black}}{}}
\usepackage[hidelinks]{hyperref}
\hypersetup{pdftitle={Port-based teleportation and quantum parallelization},
  pdfsubject={Construction and proof of a shallow port-based teleportation protocol}}
\newtheorem{theorem}{Theorem}
\newtheorem{lemma}[theorem]{Lemma}
\newtheorem{corollary}[theorem]{Corollary}
\newtheorem{claim}{Claim}[section]
\newtheorem{fact}{Fact}[section]
\newcommand{\ket}[1]{\lvert #1\rangle}
\newcommand{\bra}[1]{\langle #1\rvert}
\newcommand{\braket}[2]{\langle #1 | #2 \rangle}
\newcommand{\Tr}{\operatorname{Tr}}
\newcommand{\supp}{\operatorname{supp}}
\newcommand{\poly}{\operatorname{poly}}
\newcommand{\polylog}{\operatorname{polylog}}
\newcommand{\Swap}{\operatorname{SWAP}}
\newcommand{\id}{\operatorname{id}}

\newcommand{\EPR}{\mathrm{EPR}}

\title{A simple proof of unitary parallelization via port-based teleportation}

\author{%
\begin{tabular}{c@{\qquad}c}
Anurag Anshu & Quynh T. Nguyen \\[0.2em]
\normalsize Harvard University & \normalsize UC Berkeley
\end{tabular}}

\date{\today}

\makeatletter

\renewcommand{\maketitle}{%
  \begin{center}
    {\LARGE\@title\par}\vspace{0.7em}
    {\large\@author\par}\vspace{0.5em}
    {\@date\par}
  \end{center}\vspace{0.5em}}
\makeatother

\begin{document}

\maketitle

\begin{abstract}
Port-based teleportation (PBT) is a variant of quantum teleportation that does not require corrections on the teleported state: starting with $N$ shared EPR states, Alice can teleport a qudit to Bob by performing a measurement on her system and informing Bob which of his $N$ `ports' to retrieve the state, up to infidelity decaying with $N$. Existing efficient quantum circuits of PBT rely on representation theory tools such as the Schur transform. In this note, we report an elementary efficient circuit for PBT, which has depth $\operatorname{polylog}(N,d)$. As a corollary, this gives a simple proof that all $n$-qubit unitaries can be parallelized to $\operatorname{poly}(n)$ depth, recovering a recent result due to Nehoran--Slote--Yuen and OpenAI.
\end{abstract}

\section{Introduction}

One of the fundamental results that we learn in complexity theory courses is that any boolean function on n bits can be parallelized to $\poly(n)$ depth, at the expense of making the width exponentially large. Surprisingly, the quantum analogue of this question -- parallelizing an arbitrary unitary on $n$ qubits with a similar trade-off -- is highly non-trivial and was answered only in the recent works \cite{NSY,OAI}. An exact implementation was given in \cite{cao2026midcircuitmeasurementsfeedforwardimplement}. The goal of our note is to give a simple proof of this result via a connection to port-based teleportation (PBT)~\cite{IH}. This connection - that low depth PBT protocol would imply parallelizing a unitary-  was made explicit first in \cite{strelchuk2026powerconstantdepthquantumcircuits}.

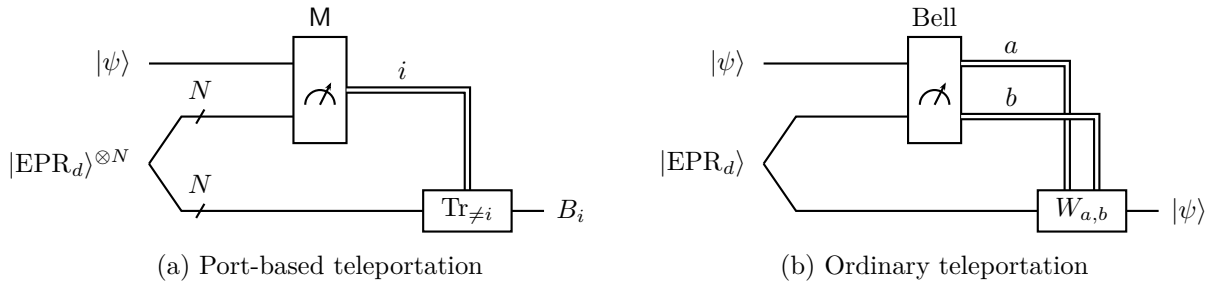
\begin{figure}[!htb]
\centering
\begin{minipage}[t]{0.49\linewidth}
\centering
\begin{tikzpicture}[x=0.78cm,y=0.78cm,thick,draw=black,
  every node/.style={font=\small,text=black},
  classical/.style={double,double distance=1.2pt}]
  % Shared EPR pairs; the two slashed wires each carry N qudits.
  \draw (0.35,-0.2) -- (0.9,0.6) -- (2.8,0.6);
  \draw (0.35,-0.2) -- (0.9,-1.0) -- (5.0,-1.0);
  \node[anchor=east] at (0.23,-0.2) {$\ket{\EPR_d}^{\otimes N}$};
  \draw (1.15,0.48) -- (1.29,0.72);
  \draw (1.15,-1.12) -- (1.29,-0.88);
  \node[above] at (1.22,0.72) {$N$};
  \node[above] at (1.22,-0.88) {$N$};
  % Alice's collective measurement.
  \draw (0.35,1.5) -- (2.8,1.5);
  \node[anchor=east] at (0.23,1.5) {$\ket\psi$};
  \draw[fill=white] (2.8,0.15) rectangle (3.7,1.95);
  \node[above] at (3.25,1.95) {$\mathsf M$};
  \draw (3.00,0.80) arc[start angle=180,end angle=0,radius=0.25];
  \draw[-{Stealth[length=3pt]}] (3.25,0.80) -- (3.44,1.18);
  % The classical outcome controls which Bob port is retained.
  \draw[classical] (3.7,1.05) -- (5.75,1.05) -- (5.75,-0.65);
  \node[above] at (4.65,1.05) {$i$};
  \draw[fill=white] (5.0,-1.35) rectangle (6.5,-0.65);
  \node at (5.75,-1.0) {$\Tr_{\ne i}$};
  \draw (6.5,-1.0) -- (7.05,-1.0);
  \node[anchor=west] at (7.08,-1.0) {$B_i$};
  \node at (3.25,-1.95) {(a) Port-based teleportation};
\end{tikzpicture}
\end{minipage}\hfill
\begin{minipage}[t]{0.49\linewidth}
\centering
\begin{tikzpicture}[x=0.78cm,y=0.78cm,thick,draw=black,
  every node/.style={font=\small,text=black},
  classical/.style={double,double distance=1.2pt}]
  % A single shared EPR pair.
  \draw (0.35,-0.2) -- (0.9,0.6) -- (2.8,0.6);
  \draw (0.35,-0.2) -- (0.9,-1.0) -- (5.0,-1.0);
  \node[anchor=east] at (0.23,-0.2) {$\ket{\EPR_d}$};
  % Alice's Bell measurement.
  \draw (0.35,1.5) -- (2.8,1.5);
  \node[anchor=east] at (0.23,1.5) {$\ket\psi$};
  \draw[fill=white] (2.8,0.15) rectangle (3.7,1.95);
  \node[above] at (3.25,1.95) {Bell};
  \draw (3.00,0.80) arc[start angle=180,end angle=0,radius=0.25];
  \draw[-{Stealth[length=3pt]}] (3.25,0.80) -- (3.44,1.18);
  % Both classical outcomes determine the Pauli correction.
  \draw[classical] (3.7,1.5) -- (5.50,1.5) -- (5.50,-0.65);
  \draw[classical] (3.7,0.6) -- (6.00,0.6) -- (6.00,-0.65);
  \node[above] at (4.55,1.5) {$a$};
  \node[above] at (4.55,0.6) {$b$};
  \draw[fill=white] (5.0,-1.35) rectangle (6.5,-0.65);
  \node at (5.75,-1.0) {$W_{a,b}$};
  \draw (6.5,-1.0) -- (7.05,-1.0);
  \node[anchor=west] at (7.08,-1.0) {$\ket\psi$};
  \node at (3.25,-1.95) {(b) Ordinary teleportation};
\end{tikzpicture}
\end{minipage}
\caption{PBT replaces Bob's outcome-dependent Pauli
correction $W_{a,b}$ by port selection: $\Tr_{\ne k}$ discards all ports
except $B_k$.}
\label{fig:teleportation-comparison}
\end{figure}

Ordinary teleportation uses one shared EPR pair: Alice performs a Bell-basis
measurement and Bob applies an outcome-dependent correction. In
port-based teleportation, Alice instead measures the input jointly with
her halves of $N$ shared pairs and reports a port label $i$. Bob keeps
$B_i$ and discards the other ports, with no correction on the retained
state (Figure~\ref{fig:teleportation-comparison}).
PBT has applications in universal programmable processors, non-local computation, AdS/CFT, cryptography, etc.

We give an elementary efficient PBT circuit with polylogarithmic depth in the number of ports and local dimension. In contrast, 
existing efficient circuits have $\poly(N,d)$ depth and are arguably more complex due to their heavy reliance on representation theory~\cite{FTH,GBO,wills2024efficient}\footnote{There are multiple variants of PBT. Here we consider
\emph{deterministic inexact PBT with EPR states}.}.

\begin{theorem}[Polylogarithmic-depth PBT]
\label{thm:main}
For every positive multiple $N$ of $d$ and $0<\delta<1$, there exists a circuit of depth
$\polylog(Nd/\delta)$ and width $\poly(N,d,\log(1/\delta))$ that implements
PBT with $N$ ports and local dimension $d$, achieving diamond-norm error
$2(d^2-1)/N+\delta$.
\end{theorem}

The absence of corrections in PBT lets us reduce the problem of implementing a unitary to the problem of preparing a state. Replacing each shared EPR pair by the Choi state $(I\otimes U)\ket{\EPR_d}$ makes the output approximately $U\ket\psi$.
These $N$ independent Choi states can be prepared in parallel using standard procedures, e.g.~\cite[Theorem~1]{ZLY}. Together, we obtain a polynomial-depth
implementation of any $n$-qubit unitary.

\begin{corollary}[Unitary parallelization]\label{cor:unitary}
For every $n$-qubit unitary $U$ and $0<\varepsilon<1$, there is a circuit of depth $\poly(n, \log(1/\varepsilon))$ and width
$2^{O(n)}\poly(1/\varepsilon)$
that implements $U$ to diamond-norm error $\varepsilon$.
\end{corollary}

\noindent 

\textbf{AI Disclosure:} The connection to port-based teleportation (already made in the prior work \cite{strelchuk2026powerconstantdepthquantumcircuits}) and low-depth eigenvalue extraction was suggested to ChatGPT Astra by the human authors as a starting point. In a few prompts, ChatGPT Astra came up with the protocol in this note, in particular the remarkable single reflection teleportation protocol. The authors have improved the manuscript for better human understanding and take the responsibility of correctness.   

\section{Teleportation from a single reflection}\label{sec:teleport}
In this section we show that a single reflection operator on Alice's registers suffices to implement PBT. The implementation of the reflection will be described in Section~\ref{sec:swaps}.

\subsection{Preliminaries}

\paragraph{Circuit model.} We use one- and two-qubit gates with arbitrary rotation angles,
unrestricted circuit connectivity, and zero-initialized ancillas.
\paragraph{PBT notations.}
Let $N\ge1$.
For an input state $\ket\psi$ to be teleported, define
\begin{equation}\label{eq:initial}
 \ket{\Psi}=\ket\psi_C\bigotimes_{i=1}^N\ket{\EPR_d}_{A_iB_i}.
\end{equation}
Let 
\[ S=\sum_{i=1}^N \ket{\EPR_d}\bra{\EPR_d}_{CA_i},
\]
as introduced by \cite{IH}.
\begin{claim}[Approximate eigenstate]\label{cl:variance}
It holds that
\begin{equation}\label{eq:variance}
 \left\|\left(\frac{d^2}{N}S-I\right)\ket{\Psi}\right\|
       =\sqrt{\frac{d^2-1}{N}}.
\end{equation}
\end{claim}
\begin{proof}
\begingroup
\color{black}
Put $P_i=\ket{\EPR_d}\bra{\EPR_d}_{CA_i}$. Use the standard EPR-state contraction identities $\Tr_{A_i}P_i
 = \frac 1d \id_C$, $\Tr_{A_iA_j}(P_iP_j) = \frac 1 {d^2} \id_C $ ($i\ne j$), and summing over $i$ and $j$ gives
\endgroup
\[
 \bra{\Psi}S\ket{\Psi}=\frac N{d^2},\qquad
 \bra{\Psi}S^2\ket{\Psi}=\frac N{d^2}+\frac{N(N-1)}{d^4}.
\]
Expanding the $\ell_2$ norm proves the claim.
\end{proof}

\subsection{Protocol}

\paragraph{Motivation:} In standard PBT, Alice performs the
pretty good measurement
\[
 E_i=S^{-1/2}\ket{\EPR_d}\bra{\EPR_d}_{CA_i}S^{-1/2}
      ,\qquad i=1,\ldots,N,
\]
where the inverse is taken on $\supp S$, and Bob keeps the port $B_i$ specified by her outcome. The algorithms of~\cite{GBO,FTH} in essence implement certain Naimark dilations of this PGM constructed from mixed Schur-Weyl duality theory.

We can define an abstract Naimark dilation as follows. For each $i$, enlarge the Hilbert space $C A_i$ and choose a normalized `flag state' $\ket f$ supported entirely on the added levels of both registers. This flag will indicate which port to keep on Bob's side.
The map $L=\sum_{i=1}^N\ket f\bra{\EPR_d}_{CA_i}$
then satisfies $L^\dagger L=S$ on Alice's data subspace because distinct
flag locations are orthogonal. On the support of $S$, the isometry \(V=LS^{-1/2}\), together with projective measurement of the flag, gives a Naimark dilation of the PGM. Thus $LS^{-1/2}$ realizes the PGM on
$\supp S$. (Handling the kernel of $S$ is simple, e.g., by outputting a random port.)

The observation is that the approximate eigenstate property in Claim~\ref{cl:variance}
motivates replacing the dilation $LS^{-1/2}$ by the simpler operator $L \cdot d/\sqrt N$,
giving the target state (see Section 6.2, \cite{strelchuk2026powerconstantdepthquantumcircuits}):
\begin{equation}\label{eq:ideal}
 \ket{\Omega}=\textcolor{black}{\frac d{\sqrt N}L\ket{\Psi}=}
 \frac1{\sqrt N}\sum_{i=1}^N
   \ket f_{CA_i}\ket\psi_{B_i}
   \bigotimes_{j\ne i}\ket{\EPR_d}_{A_jB_j}.
\end{equation}
Replacing a complicated operator such as $LS^{-1/2}$ with a simpler POVM with scalar normalization finds its roots in several prior works, such as \cite[Page 8]{WinterMassar2001}. It is readily verified that the states $\ket{\Psi}$ and $\ket{\Omega}$ are orthogonal unit vectors. Once we have $\ket{\Omega}$, Alice simply reads the flag to obtain a port label $i$, and Bob keeps $B_i$. 

Our protocol
approximately prepares this state using the reflection opearator $R_f$
defined below.

\begin{fact}\label{fact:reflection-transfer}
If a reflection $R$ fixes $\ket{\Psi}+\ket{\Omega}$ and negates
$\ket{\Psi}-\ket{\Omega}$, then $R\ket{\Psi}=\ket{\Omega}$.
\end{fact}
To construct such a reflection, we take a superposition over the EPR state and  its flag:
for $\mu>0$, define
\begin{equation}\label{eq:flag-reflection}
 \ket\beta=\ket{\EPR_d}+\sqrt\mu\,\ket f,
 \qquad S_f=\sum_{i=1}^N\ket\beta\bra\beta_{CA_i}.
\end{equation}
We choose $\mu$ to equalize the contractions of $\bra\beta_{CA_i}$
with $\ket{\Psi}$ and $\ket{\Omega}$. Claim~\ref{cl:padding} below then puts their difference
$\ket{\Psi} - \ket{\Omega}$ in $\ker S_f$ and their sum $\ket{\Psi} + \ket{\Omega}$ close to $\supp S_f$.
The support reflection $R_f=2\Pi_{\supp S_f}-I$ therefore has the signs required by Fact~\ref{fact:reflection-transfer},
up to certain error.

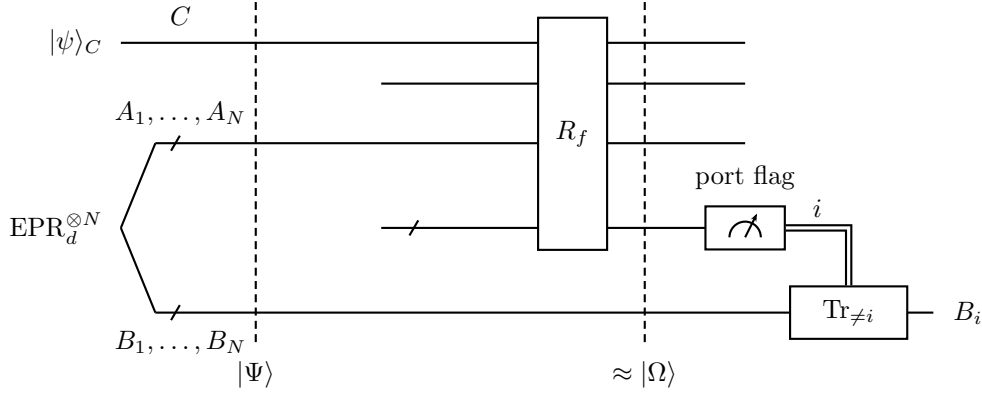
\begin{figure}[!htb]
\centering
\begingroup
\setlength{\fboxsep}{9pt}
\fbox{%
\begin{minipage}{\dimexpr\linewidth-2\fboxsep-2\fboxrule\relax}
\noindent\textbf{Protocol $\mathcal T_f$.}
Alice applies $R_f=2\Pi_{\supp S_f}-I$ and
reads the flag registers to obtain a port label $i$, or $0$ if none is flagged.
Bob keeps $B_i$. If $i=0$, Bob outputs $B_1$.

\medskip
\centering
\begin{tikzpicture}[x=0.83cm,y=0.83cm,thick,
  every node/.style={font=\small},
  gate/.style={draw,fill=white,minimum height=0.57cm,inner sep=5pt},
  classical/.style={double,double distance=1.2pt}]
  % The angled EPR branches match Figure 1.
  \draw (0.35,2.2) -- (10.3,2.2);
  \node[anchor=east] at (0.23,2.2) {$\ket\psi_C$};
  \draw (0.35,-0.75) -- (0.9,0.6);
  \draw (0.35,-0.75) -- (0.9,-2.1);
  \draw (0.9,0.6) -- (10.3,0.6);
  \draw (0.9,-2.1) -- (13.3,-2.1);
  \node[anchor=east] at (0.23,-0.75) {$\EPR_d^{\otimes N}$};
  \node[above] at (1.3,0.72) {$A_1,\ldots,A_N$};
  \node[below] at (1.3,-2.22) {$B_1,\ldots,B_N$};
  \draw (1.15,0.48) -- (1.29,0.72);
  \draw (1.15,-2.22) -- (1.29,-1.98);
  \draw[densely dashed] (2.5,2.85) -- (2.5,-2.65);
  \node[below] at (2.5,-2.7) {$\ket{\Psi}$};
  % Unlabeled wires indicate dimension extensions of Alice's registers.
  \draw (4.5,1.55) -- (10.3,1.55);
  \draw (4.5,-0.75) -- (7.0,-0.75);
  \draw (7.0,-0.75) -- (10.3,-0.75);
  \draw (4.95,-0.87) -- (5.09,-0.63);
  \node[above] at (1.3,2.32) {$C$};
  % The reflection acts on the enlarged Alice registers; Bob bypasses it.
  \draw[fill=white] (7.0,-1.1) rectangle (8.1,2.6);
  \node at (7.55,0.75) {$R_f$};
  \draw[densely dashed] (8.7,2.85) -- (8.7,-2.65);
  \node[below] at (8.7,-2.7) {$\approx\ket{\Omega}$};
  % Flag readout controls Bob's port selection.
  \node[gate,minimum width=1.05cm] (meter) at (10.3,-0.75) {};
  \draw (10.05,-0.90) arc[start angle=180,end angle=0,radius=0.25];
  \draw[-{Stealth[length=3pt]}] (10.3,-0.90) -- (10.49,-0.52);
  \node[above] at (10.3,-0.30) {port flag};
  \node[above] at (11.45,-0.75) {$i$};
  \node[gate,minimum width=1.55cm,text=black] (keep) at (11.95,-2.1)
      {$\Tr_{\neq i}$};
  \draw[classical] (meter.east) -- (11.95,-0.75) -- (keep.north);
  \node[anchor=west,text=black] at (13.45,-2.1) {$B_i$};
\end{tikzpicture}
\caption{The reflection $R_f$ approximately maps $\ket{\Psi}$
to $\ket{\Omega}$. The added wires represent the enlarged dimensions of $C$ and $A_1,\ldots,A_N$. The flag selects Bob's output port.}
\label{fig:pbt}
\end{minipage}%
}
\endgroup
\end{figure}

\begin{lemma}[PBT from a reflection]\label{lem:reflection-pbt}
Choosing $\mu=N/d^2$, the protocol in Figure~\ref{fig:pbt} implements a channel $\mathcal T_f$
satisfying $\|\mathcal T_f-\id\|_\diamond\le 2(d^2-1)/N$ for every
$N\ge1$.
\end{lemma}

\subsection{Proof of Lemma~\ref{lem:reflection-pbt}}

\begin{claim}\label{cl:unmatched-contractions}
For every $\mu>0$,
\[
 S_f(\ket{\Psi}\pm\ket{\Omega})
 =\left(1\pm\frac{d\sqrt\mu}{\sqrt N}\right)
   \left(S\ket{\Psi}+\frac{\sqrt{\mu N}}d\ket{\Omega}\right).
\]
\end{claim}
\begin{proof}
Orthogonality of the data and flag subspaces gives
\begin{align*}
\bra\beta_{CA_i}\ket{\Psi}\pm\bra\beta_{CA_i}\ket{\Omega} = \braket{\EPR_d}{\Psi}\pm \sqrt{\mu}\braket{f}{\Omega} =\left(\frac1d\pm\frac{\sqrt\mu}{\sqrt N}\right)
   \ket\psi_{B_i}\bigotimes_{j\ne i}\ket{\EPR_d}_{A_jB_j}.
\end{align*}
Bringing in $\ket\beta_{CA_i}$ on both sides and summing over $i$ proves the claim.
\end{proof}

\begin{claim}[Choice of flag weight]\label{cl:padding}
Choosing $\mu=N/d^2$ gives
\begin{equation}\label{eq:balance}
 \begin{aligned}
 S_f(\ket{\Psi}-\ket{\Omega})&=0,\\
 \left\|\left(\frac{S_f}{2\mu}-I\right)
       (\ket{\Psi}+\ket{\Omega})\right\|^2&=\frac{d^2-1}{N}.
 \end{aligned}
\end{equation}
\end{claim}
\begin{proof}
Claim~\ref{cl:unmatched-contractions} gives the first identity and
\[
 \left(\frac{S_f}{2\mu}-I\right)(\ket{\Psi}+\ket{\Omega})
 =\left(\frac S\mu-I\right)\ket{\Psi}.
\]
We conclude by applying Claim~\ref{cl:variance}.
\end{proof}

Now, we are in a position to prove Lemma~\ref{lem:reflection-pbt}.

\begin{proof}[Proof of Lemma~\ref{lem:reflection-pbt}]
Take $\ket\psi_C$ maximally entangled with a reference system.
By Claim~\ref{cl:padding},
$\Pi_{\ker S_f}(\ket{\Psi}-\ket{\Omega})=\ket{\Psi}-\ket{\Omega}$, so
\[
 R_f\ket{\Psi}-\ket{\Omega}
 =-\Pi_{\ker S_f}(\ket{\Psi}+\ket{\Omega})
 =\Pi_{\ker S_f}\left(\frac{S_f}{2\mu}-I\right)(\ket{\Psi}+\ket{\Omega}).
\]
The same claim therefore gives
\begin{equation}\label{eq:idealcoherent}
 \|R_f\ket{\Psi}-\ket{\Omega}\|^2\le\frac{d^2-1}{N}.
\end{equation}

Reading the flag and selecting $B_i$ on $\ket{\Omega}$ gives the ideal
EPR state on the output and reference. Thus~\eqref{eq:idealcoherent}
bounds the entanglement infidelity of the protocol $\mathcal T_f$.
Since $\mathcal T_f$ is unitarily covariant, the standard relation
between entanglement infidelity and diamond error~\cite[Lemma~D.2 and Corollary~3.5]{CLMSSW} gives
\(
 \|\mathcal T_f-\id\|_\diamond
 \le\frac{2(d^2-1)}{N}.
\)
\end{proof}

\section{Implementing the reflection}\label{sec:swaps}

\begin{figure}[!htb]
\centering
\begin{tikzpicture}[x=0.91cm,y=0.87cm,thick,
  every node/.style={font=\small},
  gate/.style={draw,fill=white,minimum height=0.55cm,inner sep=5pt}]
  \node[anchor=west,font=\small\bfseries] at (-0.25,0.8)
     {Reflection $R_f$};
  \draw (0,0) -- (12,0);
  \draw (0,-1) -- (12,-1);
  \node[anchor=east] at (-0.1,0) {$\ket\varphi$};
  \node[anchor=east] at (-0.1,-1) {$\ket0$};
  \node[gate] at (0.75,0) {$E$};
  \draw[fill=white] (1.65,-1.33) rectangle (4.0,0.33);
  \node[align=center] at (2.825,-0.5)
     {extract\\$k_N,k_M$};
  \node[gate] at (5.85,-1) {phase};
  \draw[fill=white] (7.15,-1.33) rectangle (9.5,0.33);
  \node[align=center] at (8.325,-0.5)
     {uncompute\\extraction};
  \node[gate] at (10.45,0) {$E^\dagger$};
  \node[anchor=west] at (12.1,0) {$R_f\ket\varphi$};
  \node[anchor=west] at (12.1,-1) {$\ket0$};
\end{tikzpicture}
\caption{Circuit for $R_f$. Here \(|\varphi\rangle\) is a state of Alice’s enlarged registers \(CA_1\cdots A_N\) (see Figure~\ref{fig:pbt}). The phase is $+1$ when
$N-\binom{D-1}{2}+k_N-k_M>0$ and $-1$ otherwise. The eigenvalue extraction is implemented with random SWAP tests.}
\label{fig:reflection}
\end{figure}
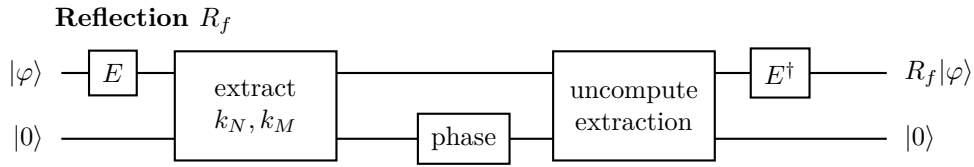

Lemma~\ref{lem:reflection-pbt} reduces PBT to implementing $R_f$.
In this section, we choose a maximally entangled flag so that $\ket\beta$ is proportional
to an EPR state in a larger dimension. This choice enables a polylogarithmic-depth circuit
for the reflection (Figure~\ref{fig:reflection}).

Fix $\mu=N/d^2$ as in Section~\ref{sec:teleport}. Assuming $d\mid N$,
enlarge Alice's registers to dimension $D=d+N/d$ and choose
\begin{equation}
   \ket f=\sqrt{\frac dN}\sum_{a=d+1}^{D}\ket{a,a},
 \qquad
 \ket\beta=\frac1{\sqrt d}\sum_{a=1}^{D}\ket{a,a}
 =\sqrt{\frac Dd}\ket{\EPR_D}.
\end{equation}
The operator $S_f$ is then proportional to the sum of EPR projectors
\begin{equation}\label{eq:sum-of-eprs}
 S_f=\frac{D}{d}\sum_{i=1}^N\ket{\EPR_D}\bra{\EPR_D}_{CA_i}.
\end{equation}
Above, we slightly abuse notation to use $C,A_i$ to denote the enlarged registers of dimension $D$.
\begin{lemma}[Reflection in polylog depth]\label{lem:reflection-circuit}
For $0<\delta<1$, there exists a circuit that implements
$R_f$ to diamond error at most $\delta$, with depth $\polylog(Nd/\delta)$
and width $\poly(N,d,\log(1/\delta))$.
\end{lemma}

The key idea to implement $R_f$ is to express $S_f$ as a sum of swap operators in an appropriate basis. Extracting the eigenvalues of $S_f$ then reduces to performing random SWAP tests, which can be done in polylogarithmic depth.

\subsection{From EPR projectors to sums of swaps}\label{sec:antisymmetric}

First, we express $S_f$ as sum of swap operators under a change of basis on the (enlarged) input register $C$, which is shared by all $\ket \beta_{C A_i}$. This is done via an isometry $E$ from register $C$ onto the antisymmetric subspace of $D-1$ registers $C_1,\ldots,C_{D-1}$ of dimension $D$, defined as follows. For each input basis label $j\in\{1,\ldots,D\}$, let $a_1<\cdots<a_{D-1}$ list
$\{1,\ldots,D\}\setminus\{j\}$ and define

\begin{equation}\label{eq:encode}
 E\ket j_{C}=\frac{(-1)^{j-1}}{\sqrt{(D-1)!}}
 \sum_{\pi\in S_{D-1}}\operatorname{sgn}(\pi)
 \ket{a_{\pi(1)},\ldots,a_{\pi(D-1)}}_{C_1\cdots C_{D-1}}.
\end{equation}
It is a standard fact that $\{E\ket j\}_{j=1}^D$ form an orthonormal basis of the antisymmetric subspace. It is also not hard to exactly implement $E$ in $\polylog(D)$ depth~\cite{liu2026low}, we describe this in Claim~\ref{cl:encoding-circuit} for completeness.

The motivation comes from the partial transpose identity
$\bigl(\ket{\EPR_D}\bra{\EPR_D}_{CA_i}\bigr)^{T_C}
  \propto \Swap_{CA_i}$.
However, partial transpose is not a physical operation. Instead, $E$ allows us to realize the same effect via the following identity.
\begin{fact}\label{fact:antisymmetric-operator-identity} For every $D\times D$ operator $X$, it holds that $
 E^\dagger\left(\sum_{t=1}^{D-1}X_{C_t}\right)E
 =\Tr(X)I-X^T.$
\end{fact}
This identity follows from elementary algebra or can be derived from~\cite[Equation 9]{miyazaki2019complex}.
\begin{claim}\label{cl:encoded-swap} It holds that
\begin{equation}
  E^\dagger\left( \sum_{i=1}^N \sum_{t=1}^{D-1}\operatorname{SWAP}_{C_t A_i}\right)E
 =NI-dS_f.
\end{equation}
\end{claim}
\begin{proof}
Expanding each swap and applying Fact~\ref{fact:antisymmetric-operator-identity}
with $X=\ket b\bra a$ gives
\[
\begin{aligned}
 E^\dagger\left(\sum_{t=1}^{D-1}\Swap_{C_t A_i}\right)E
 &=\sum_{a,b=1}^{D}
   \bigl(\delta_{ab}I_C-(\ket a\bra b)_C\bigr)
   \otimes(\ket a\bra b)_{A_i}\\
 &=I_{CA_i}-\sum_{a,b=1}^{D}(\ket{a,a}\bra{b,b})_{CA_i}
 =I_{CA_i}-D\ket{\EPR_D}\bra{\EPR_D}_{CA_i}.
\end{aligned}
\]
Summing over $i$ gives the claimed identity.
\end{proof}

We have successfully expressed $S_f$ as a sum of pairwise swaps among the registers. However, these swap operators are not quite convenient enough to implement $R_f$ in the next section. Instead, we massage the above into a more convenient swap decomposition below.

\begin{samepage}
\begin{claim}[Swap decomposition]\label{cl:swap-identity} Let $M=N+D-1$, let $K_M$ be the sum of all pairwise swaps among
$C_1,\ldots,C_{D-1},A_1,\ldots,A_N$, and let $K_N$ be the sum of swaps among $A_1,\ldots,A_N$.
It holds that
\begin{equation}\label{eq:swapidentity}
 E^\dagger \left[\left(N-\binom{D-1}{2}\right)I+K_N-K_M\right]E=dS_f.
\end{equation}

\end{claim}
\end{samepage}
\begin{proof} Decompose $K_M$ into three types of swaps: $K_N$, those within $C_1\ldots C_{D-1}$, and those across $A_1\ldots A_N$ and $C_1\ldots C_{D-1}$. Denote the last type by $K_\mathrm{cross}$. Swaps within the antisymmetric subspace of $C_1\ldots C_{D-1}$ act as $-I$, while $K_N$ does not overlap with $E$, so we have

\begin{equation}
  \label{eq:K_M expansion}
  E^\dagger K_M E
 = K_N
  - \binom{D-1}{2}I + E^\dagger K_\mathrm{cross} E.
\end{equation}
Substituting Claim~\ref{cl:encoded-swap} and rearranging concludes the claim.
\end{proof}

Conveniently, conjugating $K_M$ by any permutation simply permutes its summands, thus $K_M$ and $K_N$ commute. Furthermore, both operators also preserve the antisymmetric subspace. Therefore, the joint eigenvalues $k_N,k_M$ determine
the phase of $R_f$: $+1$ if $N-\binom{D-1}{2}+k_N-k_M>0$ and $-1$ otherwise. We therefore extract these eigenvalues coherently, apply the phase, and uncompute to implement $R_f$.

\subsection{Extracting eigenvalues with random SWAP tests}\label{sec:estimate}

We now show how to estimate eigenvalues of $K_r$ for $r \in \{N,M\}$ using random SWAP
tests in polylogarithmic depth
(Figure~\ref{fig:swap-tests}).

\begin{claim}[Swap-sum eigenvalue extraction]\label{claim:estimate}
For any $r\geq 2$ any $0<\eta<1$, there is a unitary circuit which approximates the isometry
\[
 \ket\psi\longmapsto\ket\psi\ket k\ket0_{\rm work},
 \qquad K_r\ket\psi=k\ket\psi,
\]
to diamond error at most $\eta$.
Its depth is
$\polylog(rD/\eta)$ and its width is
$\poly(r,\log D,\log(1/\eta))$.
\end{claim}

\begin{proof} Consider a state $\ket{\psi}$ with eigenvalue $k$ of $K_r$.
  
\textbf{Random SWAP tests.} Let $B=\binom r2$. The eigenvalues of $K_r$ are integers \cite[Theorem~5.1]{VO}
and lie in $[-B,B]$ because $\|K_r\|\le B$. Since every swap
commutes with $K_r$, a SWAP test on any pair of the $r$ registers preserves the $k$-eigenspace.
Imagine performing sequentially $L$ SWAP tests on randomly chosen pairs. At any time step, an outcome $x\in\{+1,-1\}$ therefore satisfies
\begin{equation}\label{eq:swap-probability}
 \Pr[x\mid k,\ \text{previous tests}]
 =\frac1B\sum_{a<b}\left\langle\frac{I+x \mathrm{SWAP}_{ab} }2\right\rangle
 =\frac12\left(1+x\frac{k}{B}\right).
\end{equation}
The outcomes $x_1,...,x_L$ are consequently independent with mean $k/B$. Hoeffding's bound gives
\begin{equation}\label{eq:swap-failure}
 \widehat k=\left\lfloor\frac BL\sum_{j=1}^Lx_j\right\rceil,
 \qquad
 \Pr[\widehat k\ne k]\le2\exp\left(-\frac{L}{8B^2}\right).
\end{equation}
Choosing $L=\lceil8B^2\ln(16/\eta^2)\rceil = \poly(r, \log(1/\eta))$ makes the failure
probability at most $\eta^2/8$. Thus, we have shown that $L$ sequential SWAP tests on random pairs of registers suffice to estimate the eigenvalue $k$ of $K_r$ with high probability. The next step is to implement these tests in parallel.

\textbf{Parallel implementation.}
Using~\cite[Theorem~1]{ZLY}, we prepare $L$ independent copies of the
uniform pair state
\[
 \frac1{\sqrt{r(r-1)}}\sum_{a\ne b}\ket{a,b}
\]
in depth $\polylog r$. Each pair encodes a uniform transposition $\tau_j$ for $j \in [L]$. We also prepare $\ket{+}^{\otimes L}$ for the SWAP test bits. Naively, we would, for each $i \in [L]$, control on the test bit $z_j$ and apply $\mathrm{SWAP}_{\tau_j}$ before the final X-basis measurement. This would lead to an undesired $\poly(L)$ depth. The key idea is to group the controlled SWAPs into a single controlled permutation,
\begin{equation}\label{eq:swap-word}
 \pi(z,\tau)=\tau_L^{z_L}\cdots\tau_1^{z_1}.
\end{equation}
which can be computed and implemented in depth $\polylog (LrD)$ using Claim~\ref{cl:parallel-primitives} for elementary arithmetic circuits. Finally, we apply $H^{\otimes L}$ to the test bits and read out the estimate.
\begin{figure}[H]
\centering
\begin{tikzpicture}[x=0.91cm,y=0.87cm,thick,
  every node/.style={font=\small},
  gate/.style={draw,fill=white,minimum height=0.55cm,inner sep=5pt}]
  \node[anchor=west,font=\small\bfseries] at (-0.25,-2.35)
     {Parallel random SWAP tests};
  \draw (1.1,-3.15) -- (10.9,-3.15);
  \draw (1.1,-4.05) -- (10.9,-4.05);
  \draw (1.1,-4.95) -- (10.9,-4.95);
  \node[anchor=east] at (0.98,-3.15) {test bits $\ket{0^L}$};
  \node[anchor=east] at (0.98,-4.05) {pair choices $\ket{\tau_1,...,\tau_L}$};
  \node[anchor=east] at (0.98,-4.95) {data};
  \node[gate] at (2.05,-3.15) {$H^{\otimes L}$};
  \node[gate,minimum width=3.7cm] (perm) at (6.0,-4.95)
     {$U_{\pi(z,\tau)}$};
  \draw (6.0,-3.15) -- (perm.north);
  \fill (6.0,-3.15) circle (1.7pt);
  \fill (6.0,-4.05) circle (1.7pt);
  \node[above] at (4.1,-3.15) {$z$};
  \node[above] at (4.1,-4.05) {$\tau$};
  \node[gate] at (9.4,-3.15) {$H^{\otimes L}$};
  \node[anchor=west] at (11.0,-3.15) {test outcomes};
  \node[below] at (6.0,-5.36)
     {$\pi(z,\tau)=\tau_L^{z_L}\cdots\tau_1^{z_1}$};
\end{tikzpicture}
\caption{Coherent implementation of $L$ random SWAP tests.}
\label{fig:swap-tests}
\end{figure}
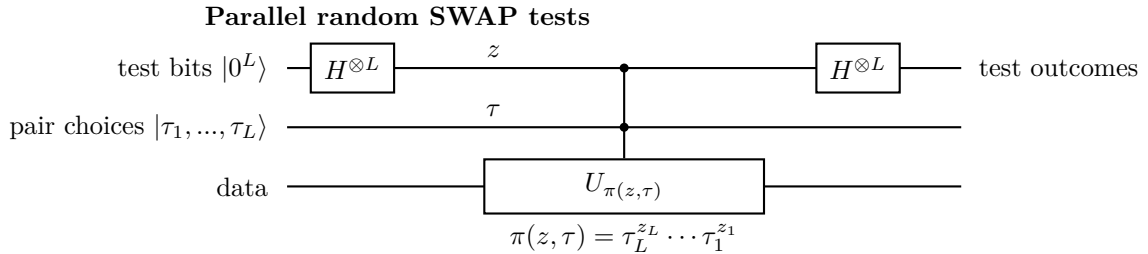

\textbf{Coherent extracting $\hat{k}$.}
Using Claim~\ref{cl:parallel-primitives} for parallel arithmetic circuits,
we compute $\widehat k$ from the test bits, copy it to an output register,
and reverse the entire computation, including the tests and pair
preparation. On a $k$-eigenstate, the Hoeffding's bound guarantees squared norm error $2\Pr[\widehat k\ne k]\le\eta^2/4$. The test circuit and the estimate computation preserve each
$K_r$-eigenspace, as do their inverses. Errors from distinct
eigenspaces are therefore orthogonal, giving $\ell_2$-norm error
at most $\eta/2$ on the output states, hence the diamond error is at most $\eta$.
\end{proof}

\subsection{Applying the support reflection}
We combine Claim~\ref{cl:swap-identity} and Claim~\ref{claim:estimate} to conclude Lemma~\ref{lem:reflection-circuit}.

\begin{proof}[Proof of Lemma~\ref{lem:reflection-circuit}]
Claim~\ref{cl:encoding-circuit} implements the isometry $E$ in depth $\polylog D$ and width $\poly D$. Next apply
Claim~\ref{claim:estimate} to $K_N$ and $K_M$, extracting $k_N,k_M$ with diamond
error at most $\delta/4$ each. Then, we use elementary arithmetic circuits from Claim~\ref{cl:parallel-primitives} to compute $s=N-\binom{D-1}{2}+\hat k_N- \hat k_M$ in depth $\polylog(MD)$ and
width $\poly(M,\log D)$. Apply phase $+1$ if $s>0$ and $-1$ otherwise,
uncompute this arithmetic, then reverse both eigenvalue extraction circuits and the
encoding $E$. This gives $R_f$ when the labels
are extracted exactly. The diamond-norm errors of $k_n,k_M$ extraction procedures and their inverses add up, so we implement $R_f$ up to diamond error $\delta$.
\end{proof}

\section{Putting the circuit together}\label{sec:resources}
We now prove Theorem~\ref{thm:main}.

\begin{proof}[Proof of Theorem~\ref{thm:main}]
We use the protocol of Section~\ref{sec:teleport}, with the EPR flag of Section~\ref{sec:swaps}, and implement $R_f$ to diamond error $\delta$ using
Lemma~\ref{lem:reflection-circuit} in depth
$\polylog(Nd/\delta)$ and width $\poly(N,d,\log(1/\delta))$. Using arithmetic circuits from Claim~\ref{cl:parallel-primitives}, Alice next identifies the flagged port $i$ (or $0$ if none; breaking tie by choosing smalles port index). Conditioned on this label, Bob exchanges his port $i$ with port $1$, doing nothing for label $0$. The output of the overall PBT protocol is Bob's port 1. Lemma~\ref{lem:reflection-pbt} and triangle inequality gives the overall diamond norm bound
$2(d^2-1)/N+\delta$.
\end{proof}

Next, we apply Theorem~\ref{thm:main} with Choi state resources to prove Corollary~\ref{cor:unitary}.

\begin{proof}[Proof of Corollary~\ref{cor:unitary}]
Apply Theorem~\ref{thm:main} with $d=2^n$,
$N=d\lceil4d/\varepsilon\rceil$, and $\delta=\varepsilon/2$,
replacing each EPR pair by the Choi state $(I\otimes U)\ket{\EPR_d}$. The $N$ Choi state copies can be prepared exactly in parallel
in $\poly(n)$ depth and $N2^{O(n)}$ width~\cite[Theorem~1]{ZLY}. Since $U^{\otimes N}$ commutes with Alice's operations and the controlled
permutation of Bob's ports, this is equivalent to PBT followed by $U$
on the selected port. The diamond error in implementing $U$ is therefore at most
$ 2(d^2-1)/N+\varepsilon/2 \le\varepsilon.$
Since $N=\poly(d,1/\varepsilon)$, the combined circuit has depth
$\poly(n,\log(1/\varepsilon))$ and width
$2^{O(n)}\poly(1/\varepsilon)$.
\end{proof}

\subsection*{Acknowledgement}

AA acknowledges support through the NSF Award Nos. 2238836, 2430375 and QCIS-FF: Quantum Computing $\&$ Information Science Faculty Fellow at Harvard University (NSF 2013303)

\printbibliography

\appendix
\small

\section{Elementary parallel circuits}\label{app:circuits}
We give the elementary arithmetic and quantum circuits used in the main text.

\begin{claim}[Parallel arithmetic and permutations]\label{cl:parallel-primitives}
The following operations admit exact circuits (with all workspace qubits reset to zero at the end)
\begin{enumerate}
\setlength{\itemsep}{2pt}
\setlength{\parskip}{0pt}
\item \textbf{Arithmetic.} Sums, comparisons, and minima of $L$ $b$-bit integers can be computed
in depth $\polylog(Lb)$ and width $\poly(L,b)$.
\item \textbf{Controlled swaps.} Given distinct $a,b\in\{1,\ldots,r\}$ and a bit $z\in\{0,1\}$,
define $\pi=(a\,b)^z$, where $(a\,b)$ swaps $a$ and $b$ and fixes all other labels.
Thus $\pi$ is the identity when $z=0$ and this swap when $z=1$.
The tuple $(\pi(1),\ldots,\pi(r))$ can be computed exactly in depth
$O(\log r)$ and width $\poly(r)$.
\item \textbf{Multiplying permutations.} The product $\pi_L\cdots\pi_1$ of $L$ permutations of $r$ objects,
each given as a list $(\pi_j(1),\ldots,\pi_j(r))$, can be computed
in depth $\polylog(rL)$ and width $\poly(r,L)$.
\item \textbf{Applying permutations.} Given a permutation as a list $(\pi(1),\ldots,\pi(r))$,
the corresponding permutation of $r$ $D$-dimensional registers can be applied
in depth $\polylog(rD)$ and width $\poly(r,\log D)$.
\end{enumerate}
\end{claim}
\begin{proof}
\textbf{Arithmetic.} Addition and comparison have logarithmic depth in the number of bits. Combining pairs in $O(\log L)$ rounds gives sums and minima.

\textbf{Controlled swaps.} Comparing each index with $a$ and $b$ and
using the control bit $z$ gives $(\pi(1),\ldots,\pi(r))$ in parallel
in depth $O(\log r)$.

\textbf{Multiplying permutations.} For permutation composition, use
$(\sigma\pi)(a)=\sigma(\pi(a))$: all $r$ values can be computed simultaneously by first copying using CNOTs. To multiply $L$ permutations, compose
consecutive pairs in parallel, then repeat on the resulting products. This takes $O(\log L)$ rounds.

\textbf{Applying permutations.} Every permutation of $r$ elements can be
routed through a fixed bitonic sorting network with $O(\log^2 r)$ layers
of switches. Given the list $(\pi(1),\ldots,\pi(r))$, compute the switch
settings by sorting a copy of the list and recording each comparison's swap decision. This takes $\polylog r$ depth. We also need to copy the switch bits to swap entire $D$-dimesional registers, giving total depth $\polylog(rD)$.
\end{proof}

% \begin{samepage}
% \begin{claim}[Uniform ranges and pairs]\label{cl:uniform-pairs}
% For every integer $r\ge1$, the state
% $r^{-1/2}\sum_{a=1}^r\ket a$ can be prepared exactly in depth
% $O(\log r)$ and width $\poly(r)$. For $r\ge2$, the same holds for
% \(
%  (r(r-1))^{-1/2}\sum_{a\ne b}\ket{a,b}
% \).
% \end{claim}
% \end{samepage}
% \begin{proof}
% First prepare the $W$ state
% $r^{-1/2}\sum_{a=1}^r\ket{0^{a-1}1\,0^{r-a}}$.
% Starting with $\ket{10...0}$, split its amplitude along a binary tree
% using two-qubit rotations, with probabilities proportional to the sizes of the two tree branches. Once we have the $W$ state, we convert $\ket{0^{a-1}1\,0^{r-a}}$ to $\ket a$ by computing parities using Claim~\ref{cl:parallel-primitives}.

% For the superposition over pairs, prepare uniform superpositions $\sum_{a=1}^r\ket a$ and $\sum_{u=1}^{r-1}\ket u$, and compute
% $b=u+\mathbf1_{u\ge a}$ using Claim~\ref{cl:parallel-primitives}. This gives uniform ordered pairs $b\ne a$.
% Erase $u$ using $u=b-\mathbf1_{b>a}$.
% \end{proof}

\begin{claim}[Antisymmetric encoding]\label{cl:encoding-circuit}
For an integer $D\ge2$, the isometry $E$ from Section~\ref{sec:antisymmetric} can be implemented exactly in depth $\polylog D$ and
width $\poly(D)$ (with all workspace reset to zero at the end).
\end{claim}
\begin{proof} Recall $E\ket j_{C}=\frac{(-1)^{j-1}}{\sqrt{(D-1)!}}
 \sum_{\pi\in S_{D-1}}\operatorname{sgn}(\pi)
 \ket{a_{\pi(1)},\ldots,a_{\pi(D-1)}}_{C_1\cdots C_{D-1}}$.

We first prepare $\sum_{\pi\in S_{D-1}} \ket{\pi(1),...,\pi(D-1)}$. For this, we begin by preparing in $\polylog D$ depth~\cite[Theorem~1]{ZLY} the uniform
superpositions $\sum_{r_k=1}^k \ket{r_k}$ for $k=1,\ldots,D-1$. There is one-to-one correspondence between the list $(r_1,\ldots,r_{D-1})$ and a permutation $\pi\in S_{D-1}$ as follows. Starting from the empty list $()$ and $k=1$, we insert $k$ at position $r_k$ of the current list. Then we identify the resulting list with $(\pi(1),...,\pi(D-1))$. Vice versa, $(r_1,...,r_{D-1})$ can be recovered from $(\pi(1),...,\pi(D-1))$ by removing entries starting from $D-1$ and recording the position $r_k$ of the removal of label $k$. Thus the correspondence is bijective. Being able to implement this correspondence in $\polylog(D)$ depth would allow us to convert $\bigotimes_{k=1}^{D-1} \big(\sum_{r_k=1}^k \ket{r_k} \big)$ into the desired $\sum_{\pi\in S_{D-1}} \ket{\pi(1),...,\pi(D-1)}$. The insertion definition of the correspondence is sequential, so we instead use the following characterization.

Define the cycle $\tau_k=(k\;k-1\;\cdots\;r_k)$, with
$\tau_k=\id$ when $r_k=k$. Then it can be verified that
\[
 \pi=\tau_2\circ\tau_3\circ\cdots\circ\tau_{D-1}.
\]
 Each permutation $\tau_k$ depends only on $r_k$, so Claim~\ref{cl:parallel-primitives}
computes their list representations and product $\pi$ in depth $\polylog D$. This gives us parallel implementation of one direction of the correspondence.
For the inverse, when $k$ was inserted, exactly $r_k-1$ smaller labels preceded it.
Later insertions add only larger labels and preserve the relative order of existing labels. Therefore,
\[
 r_k=1+\#\bigl\{a\in\{1,\ldots,k-1\}:\pi^{-1}(a)<\pi^{-1}(k)\bigr\}.
\]
Here $\pi^{-1}(a)$ is the position of label $a$.
Claim~\ref{cl:parallel-primitives} computes all these counts in parallel in depth $\polylog D$.

Next, apply the phase $(-1)^{j-1}\operatorname{sgn}(\pi)$ by computing the parity
of $j-1$ and the number of out-of-order pairs in the list representation of $\pi$.

Finally, replace each label $\ell$ in $(\pi(1),...,\pi(D-1))$ by $\ell+\mathbf1_{\ell\ge j}$.
At this stage we have $$\ket{j} \frac{(-1)^{j-1}}{\sqrt{(D-1)!}}
 \sum_{\pi\in S_{D-1}}\operatorname{sgn}(\pi)
 \ket{a_{\pi(1)},\ldots,a_{\pi(D-1)}}_{C_1\cdots C_{D-1}}.$$
Erase $j$ using
$j=D(D+1)/2-\sum_t a_{\pi(t)}$.
Again, arithmetic circuits from Claim~\ref{cl:parallel-primitives} give
$\polylog D$ depth.
\end{proof}

\end{document}